\documentclass[epjc3,smallextended,final,twocolumn]{svjour3}

\usepackage{cite} 
\usepackage[english]{babel} 
\usepackage[utf8]{inputenc}
\usepackage{latexsym} 
\usepackage{amsmath} 
\usepackage{amsfonts} 
\usepackage{amssymb} 
\usepackage{textcomp} 
\usepackage{graphicx}  
\usepackage{enumerate} 
\usepackage{array} 
\usepackage{mathrsfs}
\usepackage{bm,bbm}
\usepackage{bbold}  
\usepackage{multirow}
\usepackage{fancyref}
\usepackage[breaklinks]{hyperref}
\usepackage{color}
\newcommand{\Eq}[1]{(\ref{#1})}

\newcommand{\be}{\begin{equation}}
\newcommand{\ee}{\end{equation}}
\newcommand{\ba}{\begin{eqnarray}}
\newcommand{\ea}{\end{eqnarray}}
\newcommand{\bs}{\begin{subequations}}
\newcommand{\es}{\end{subequations}}
\def\com{\color{magenta}}
\def\cob{\color{blue}}
\newcommand{\rmd}{{\rm d}}
\newcommand{\rmi}{{\rm i}}
\newcommand{\rme}{{\rm e}}

\newcommand{\arX}[1]{\href{http://arxiv.org/abs/#1}{{\ttfamily\com arXiv:#1}}}
\newcommand{\oarX}[1]{\href{http://arxiv.org/abs/#1}{{\ttfamily\com arXiv:#1}}}

\newcommand{\doin}[6]{\href{http://dx.doi.org/#1}{\cob  #2 #3 {\bf #4}, #5 (#6)}}
\newcommand{\doinn}[5]{\href{http://dx.doi.org/#1}{\cob  #2 {\bf #3}, #4 (#5)}}
\newcommand{\doij}[5]{\href{http://dx.doi.org/#1}{\cob  #2 {\bf #3}, #4 (#5)}}

\newcommand{\tia}[1]{#1.}
\newcommand{\tiap}[1]{#1}

\def\lp{\ell_{\rm Pl}}

\def\p{\partial}
\def\a{\alpha}
\def\b{\beta}

\def\de{\delta}
\def\g{\gamma}

\def\la{\lambda}

\def\om{\omega}

\def\cL{\mathcal{L}}

\def\cF{\mathcal{F}}

\renewcommand{\dh}{d_\textsc{h}}
\newcommand{\ds}{d_\textsc{s}}
\newcommand{\dw}{d_\textsc{w}}
\newcommand{\Pl}{{\text{\tiny Pl}}}
\def\lp{\ell_\Pl}
\def\tp{t_\Pl}
\def\mpl{m_\Pl}

\journalname{Eur. Phys. J. C}

\date{March 9, 2016}

\begin{document}\sloppy 

\title{Lorentz violations in multifractal spacetimes}

\author{Gianluca Calcagni\thanksref{addr1,e1}}
\thankstext{e1}{e-mail: calcagni@iem.cfmac.csic.es}
\institute{Instituto de Estructura de la Materia, CSIC, Serrano 121, 28006 Madrid, Spain\label{addr1}}

\maketitle

\begin{abstract}
Using the recent observation of gravitational waves (GW) produced by a black-hole merger, we place a lower bound on the energy above which a multifractal spacetime would display an anomalous geometry and, in particular, violations of Lorentz invariance. In the so-called multifractional theory with $q$-derivatives, we show that the deformation of dispersion relations is much stronger than in generic quantum-gravity approaches (including loop quantum gravity) and, contrary to the latter, present observations on GWs can place very strong bounds on the characteristic scales at which spacetime deviates from standard Minkowski. The energy at which multifractal effects should become apparent is $E_*>10^{14}\,{\rm GeV}$ (thus improving previous bounds by 12 orders of magnitude) when the exponents in the measure are fixed to their central value $1/2$. We also estimate, for the first time, the effect of logarithmic oscillations in the measure (corresponding to a discrete spacetime structure) and find that they do not change much the bounds obtained in their absence, unless the amplitude of the oscillations is fine tuned. This feature, unavailable in known quantum-gravity scenarios, may help the theory to avoid being ruled out by gamma-ray burst (GRB) observations, for which $E_*> 10^{17}\,{\rm GeV}$ or greater. 
\end{abstract}



\section{Introduction} 

Tests of Lorentz violations are among the most powerful tools by which experiments can constrain theories going beyond classical general relativity and the Standard Model of quantum interactions. In particular, the application of effective models of quantum gravity and string theory giving rise to phenomenological dispersion relations $E^2[1+O(1)(E/M)^n]=k^2$ has been severely limited by accurate bounds on time delay of photons coming from distant sources such as GRBs, highly energetic flares in active galactic nuclei, and emissions from pulsars. A recompilation of results can be found in \cite{HESS}. The main message from these searches is that, in general, the mass scale $M$ at which quantum-gravity effects modify the dispersion relation of photons is either very high ($M>M_2=10^6-10^{11}\,{\rm GeV}$ for $n=2$ and $M>M_1=10^{15}-10^{18}\,{\rm GeV}$ for $n=1$) or even larger than the Planck mass for $n=1$ (as found by the Fermi Gamma-ray Space Telescope \cite{Fermi,Vas13}). Therefore, while the frantic search for new physics in accelerators and in the sky continues, we already have strong bounds on certain classes of quantum-gravity models. These constraints come from the propagation of photons but it would be interesting to obtain independent bounds more related to the physics of massive bodies. In other words: What can gravity say about quantum gravity?

This question has received an answer recently, following the discovery of gravitational waves emitted from a black-hole merger \cite{Abb16}. It turns out that the same dispersion relations constrained by GRBs, and applicable also to gravitons, are poorly constrained by the low-frequency GWs typically produced by black holes, getting in fact a lower bound $M>10^{-4}-10^5\,{\rm eV}$ \cite{EMNan,ArCa2}. Although this provides an interesting proof of concept about the existence of independent checks on exotic dispersion relations, it is somewhat disappointing. The next question we pose is then: \emph{Is there any theory of nonstandard classical or quantum geometry that can be efficiently constrained by gravitational waves of astrophysical origin?}

The answer is in the affirmative. Multifractional theories (initiated in \cite{fra1,frc1,frc2}; see \cite{revmu} and references therein for an comprehensive review) are a proposal according to which the geometry of spacetime is characterized, in the simplest formulation, by a fundamental length scale $\ell_*$, a time scale $t_*$, and an energy scale $E_*$. The geometry changes depending on the scale of observation and has the typical features of a multifractal: a very ``irregular'' set similar to itself in any zoomed-in region and whose dimension changes with the scale \cite{trtls}. At ultra-microscopic scales $\lp<\ell\ll\ell_*$, spacetime is discrete, while at microscopic scales $\ell\sim\ell_*$ it is coarse-grained to a continuum. At macroscopic scales $\ell\gg\ell_*$, it reduces to an ordinary four-dimensional manifold.

A first motivation to consider multiscale theories is the possibility of improving the renormalizability of perturbative quantum gravity \cite{fra1,frc2}. A second reason is that, even when gravity is not directly quantized in this framework, the multifractal structure superimposed to the metric reproduces essentially the same regimes (at scales $\ell\sim\ell_*$) of several quantum gravities. In fact, whenever there is a change of dimensionality of position and/or momentum space (a phenomenon, called dimensional flow, typical of quantum gravity), the measure $q(x)$ used in multifractional theories is the most general at mesoscopic (i.e., super-Planckian) and large length scales, under the assumption of factorizability of the coordinate dependence. A small theorem recently proved this in a dynamic- and background-independent way: while dimensional flow {per se} determines, with an expansion of the effective spacetime dimension in the infrared, the form of $q(x)$, dynamics fixes the choice of the parameters in the expansion, thus giving rise to the abundant variety of dimensional flows in different approaches \cite{first}.

Extant bounds on the scales $\ell_*$, $t_*$, and $E_*$ come from quantum particle physics, in particular the muon lifetime (weak interactions), the Lamb shift in hydrogenic atoms (electrodynamics) and the value of the fine-structure constant \cite{frc13,frc12}. These bounds are just the beginning of a full comparison between the theory, which has reached a certain level of maturity, and experiments.

In this paper, we take one of the multifractional theories under better analytic control, that with $q$-derivatives, and find a dispersion relation with $n<1$ (a feature unique to this theory, as far as we know), a correction much less suppressed than in quantum gravity. Contrary to other approaches where similar dispersion relations are phenomenological (i.e., \textit{motivated} or \textit{inspired} by quantum gravity or string theory), our expression will be \textit{derived} directly from a full theory. We thus obtain, from GWs, the strongest bounds ever on the scales of the geometry, improving the independent constraints of \cite{frc12,frc13}. However, we also meet with the challenge to account for the GRB bounds on Lorentz violations. This will force us to explore a previously ignored sector of the theory. Here, we establish six results. (i) Confirming recent findings on (non)relativistic motion \cite{trtls}, we see that GWs are faster (respectively, slower) than ordinary light if the multifractional structure of the geometry is limited to the time (spatial) direction(s). (ii) The effect of an exotic geometry on dispersion relations is much stronger than in generic quantum-gravity approaches (including loop quantum gravity) and, contrary to the latter, it can be vigorously constrained by GW observations. (iii) Against naive expectations, astrophysical observations do not necessarily lead to stronger bounds than Standard-Model experiments, at least in the case of gravitational waves. (iv) However, fixing the fractional exponents in the measure to their central value, we do improve previous bounds on the scales of the measure by 12 orders of magnitude. In particular, the energy at which multifractal effects should become apparent is $E_*>10^{14}\,{\rm GeV}$. (v) The effect of logarithmic oscillations in the measure (corresponding to a discrete spacetime structure) changes the bounds obtained in their absence by no more than one order of magnitude, unless the amplitude of the oscillations is fine tuned. (vi) Point (v) may be crucial to avoid the theory being ruled out by GRB observations, as we will discuss at the end.

It may be useful to compare our framework with the better known Lorentz-violating general extension of the Standard Model \cite{CoKo}. The Standard-Model extension (SME) is an effective field-theory approach parametrizing all possible Lorentz- and CPT-violating operators that can be added to the strong and electroweak standard sectors. The main reason to be interested in these operators is that they may represent corrections coming from a fundamental theory of quantum gravity. Their effects can be constrained by an impressive battery of particle-physics experiments \cite{KoRu} and obviate the problem of detecting unobservably small Planck-scale modifications that any such theory would predict in a low-curvature approximation. This is the same spirit moving us to study multifractional theories, but with some notable differences. First, the form of Lorentz-violating operators in the SME can mimic some of the multifractional effects, but not many of them and never completely, essentially because no pre-fixed measure factors appear in the SME (question {\it 34} of \cite{revmu}). Second, the effects found here and in \cite{frc12,frc13} are not the product of an effective-field-theory approximation of a more fundamental theory: they are a direct manifestation of the underlying anomalous geometry, encoded in a fundamental action of particle interactions and gravity (these actions can be found in \cite{frc12,frc13,frc11} and are collected in questions {\it 31} and {\it 40} of \cite{revmu}). For this reason, while in the gravitational SME the fundamental theory is Lorentz invariant and symmetry breaking is spontaneous, multifractional theories break Lorentz invariance explicitly (although explicit Lorentz breaking occurs also in the nongravitational version of the SME). Moreover, the energy-momentum tensor is conserved as usual in the fundamental theory at the origin of the SME, while in our case the conservation law is heavily modified \cite{frc12,frc13,frc11}.


\section{Dispersion relations in quantum gravity}\label{qgr}

In general, the emission of gravitational waves and their wave-form strongly depend on the theory describing the astrophysical source emitting the signal.\footnote{The emission rate is calculated from the energy-momentum tensor via the quadrupole formula.} These details may or may not influence the determination of the propagation of the gravitational waves, depending on the method used. For instance, if one considers the propagation time between the source and Earth, then the constraint on the propagation speed $v$ may be affected by the physics around the emission point, and one may have to consider any modification in the quadrupole formula induced by the theory (quantum gravity, multifractal spacetimes, and so on). However, the constraint on the propagation speed $v$ by LIGO was placed by measuring the difference between two detectors in the time arrival of the wave front, in which case one can focus on the dispersion relation of the wave front (in the particle-physics language, of the graviton). In other words, even if the wave-form at the emission point is modified by theory, LIGO constraints on the propagation speed are not affected because they are obtained at Earth. Also, a binary system does not emit an isotropic wave front but, under the same assumptions (large source-observer distance and local multi-detector measurements), the only dependence from the position of the observer is in the intensity of such signal at the source, not in its propagation. 

Therefore, for the purpose of constraining the propagation speed it is sufficient to consider its dispersion relation in vacuum. This is the traditional starting point of the phenomenology of quantum gravity we will consider below, and it will be valid also for the multifractional case.

Given a dispersion relation $E^2=E^2({\bf k})$, the magnitude of the velocity of propagation of a wave front is given by the absolute value of the group velocity ${\bf v}$:
\be\label{gruve} 
v:=|{\bf v}|=\left|\frac{\rmd E}{\rmd {\bf k}}\right|\,.
\ee
In general, $v\neq \rmd E/\rmd|{\bf k}|$, unless the dispersion relation is isotropic and depends only on $k:=|{\bf k}|$. We will assume this throughout the paper, so that $v=\rmd E(k)/\rmd k$. For the usual Lorentz-invariant dispersion relation $E^2=k^2+m^2$ (we work in $c=1=\hbar$ units), in the small-mass limit one gets the difference $\varDelta v:= v-1
\simeq -m^2/(2E^2)$ between the propagation speed of the signal and the speed of light. The mass of the graviton can be constrained from the observation of GWs produced by massive binary systems \cite{Wil97}.

In string theory and ``quantum gravity'' at large, general considerations lead to the effect \cite{ACEMN} 
\be\label{devqg}
\varDelta v_\textsc{qg} \simeq - b_1\,\frac{E}{M}\,,
\ee
with unspecified constant factor $b_1=O(1)$. On the other hand, arguments concerning black-hole thermodynamics effectively describe the propagation of GWs
by a logarithmic dispersion relation $E^2\propto\ln[1+8\pi k^2/(3M^2)]$ in four dimensions \cite{Pad98,Pad99,ArCa1}, such that \cite{ArCa2}
\be\label{devbh}
\varDelta v_{{\rm nl},\textsc{lqg}} \simeq -3b_2\left(\frac{E}{M}\right)^2,
\ee
where $b_2=8\pi/9$. In loop quantum gravity, one can argue that the expected modification to the dispersion relation is of cubic order, $E \simeq k+b_2E^3/M^2$, where $b_2=O(1)$ \cite{ACAP}. This leads again to Eq.\ \Eq{devbh} but with generic $b_2$ \cite{ArCa2}.

Reference \cite{Abb16} gave the upper bound $m<1.2\times 10^{-22}\,{\rm eV}$ for the mass of the graviton, corresponding to
\be\label{devobs}
|\varDelta v| < 1.7 \times 10^{-18}\,,\qquad E=h\nu\approx 6.6\times 10^{-14}\,{\rm eV}\,,
\ee
where $\nu$ is the frequency the signal of event GW150914 is peaked at. Notice that Eqs.\ \Eq{devqg} and \Eq{devbh} are strongly suppressed for these frequencies, so that the constraint $M>10^{-4}-10^5\,{\rm eV}$ from GWs is very weak \cite{ArCa2,EMNan}. The bounds coming from GRBs are much stronger: $M>M_1$ for the linear case \Eq{devqg}, under strong pressure \cite{HESS} or nearly ruled out \cite{Fermi,Vas13}; $M>M_2$ for the quadratic loop-quantum-gravity case \Eq{devbh}. However, they do not apply to pure-gravity modifications, such as the nonlocal logarithmic model, although the effect \Eq{devbh} on gravitons is the same.


\section{Dispersion relations in multifractional spacetimes} 


\subsection{General paradigm} 

Before discussing the dynamics of these theories, we recall some basic facts about their kinematical geometric structure. The starting point is to assume dimensional flow, i.e., the spacetime dimension changes with the scale. A spacetime with such a property is called multiscale because dimensional flow requires the existence of at least one fundamental scale in the geometry. This assumption is inspired by quantum gravities: all the extant theories have been found to be characterized by dimensional flow (see \cite{revmu} for a more detailed discussion, examples and a full list of references) and it is an open question whether the latter is just a mathematical feature or, on the other hand, an observable one. In the second case, we would have a great instrument to test quantum gravity in a number of experiments. Moreover, although dimensional flow is not responsible \emph{per se} for the improvement of the renormalization of the gravitational interaction \cite{revmu}, it can contribute to it nevertheless. For these reasons, it is desirable to develop a formalism where dimensional flow is under analytic control.

This is the foundational principle of multifractional theories. Requiring spacetime to be multiscale and that one reaches the infrared as an asymptote (``slow'' dimensional flow at large scales and late times) are two very general background-independent and dynamics-independent assumptions satisfied in all quantum gravities. The surprise (proven in two theorems \cite{revmu,first}) is that they are enough to determine the general profile of the spacetime dimension, at least at mesoscopic-to-large scales. Here we recall only the main result of the second flow-equation theorem, which applies to the special case of factorizable measures. All spacetimes in $D$ topological dimensions where the Hausdorff dimension $\dh$ (roughly speaking, the scaling of volumes with their linear size) is multiscale have a measure $\rmd^Dq(x)$ with a specific form dictated by the first flow-equation theorem \cite{revmu,first}. For purely technical reasons related to the possibility to have a self-adjoint quadratic Laplace--Beltrami operator, we concentrate on factorizable measures, to which the second flow-equation theorem applies. In $D=4$, the multiscale measure is
\be
\rmd^4q(x)=\rmd q^0(t)\,\rmd q^1(x^1)\cdots \rmd q^3(x^3)\,, 
\ee
where the four profiles $q^\mu(x^\mu)$ (called geometric coordinates) depend on a hierarchy of length and time scales $\ell_n^\mu$. These geometries are called multifractional and are characterized by being multiscale and having measures and Laplacians factorizable in the coordinates. The most general form of $q^\mu(x^\mu)$ \cite{first} can be reduced to a simple one with only two length scales $\ell_*$ and $\ell_\infty$ and two time scales $t_*$ and $t_\infty$ in the hierarchy. This measure, called binomial, is all we need to get nontrivial effects of dimensional flow and encodes the anomalous scaling of correlation functions typically found in quantum gravities. In the ``isotropic'' case, all spatial directions $\mu=i=1,2,3$ have the same anomalous scaling $\a_i=\a$ and the geometric coordinates for one frequency $\om_N$ are \cite{revmu,first}
\ba
q^i(x^i) &=& x^i+\frac{\ell_*}{\a}\left|\frac{x^i}{\ell_*}\right|^\a F_\om(x^i)\,,\label{qx}\\
q^0(t) &=& t+\frac{t_*}{\a_0}\left|\frac{t}{t_*}\right|^{\a_0} F_\om(t)\,,\label{qt}
\ea
where $\a_\mu=\a_0,\a$ is limited to the range $0<\a_\mu<1$,\footnote{The range $0<\a_\mu<1$ cannot be extended. Negative values of $\a_\mu$ would lead to a problematic negative dimensionality of space and/or time, while values greater than 1 would lead to a wrong infrared limit of the measure (question \textit{08} of \cite{revmu}).} $F_\om(x)=1+A\cos(\om_N\ln|x/\lp|)+B\sin(\om_N\ln|x/\lp|)$, $A$ and $B$ are constant amplitudes, and $\om_N=2\pi\a/\ln N$ with $N=2,3,\dots$. The Planck length $\lp$ appears (quite unexpectedly, from a nontrivial connection between multifractional and noncommutative spacetimes \cite{ACOS}) at the bottom of the scale hierarchy $\ell_*\geq\ell_\infty=\lp$ of the measure. In the time direction, $\lp$ is replaced by $t_\infty=\tp$.

Exactly the same measure arises when completely forgetting about the flow-equation theorem (that relies only on having a slow dimensional flow in the infrared) and asking, instead, to build the continuum approximation of the measure of a deterministic multifractal \cite{frc1,frc2}. Very specific rules of fractal geometry give the same result \Eq{qx} and \Eq{qt} and help in the interpretation of these spacetimes. For instance, the log oscillations in \Eq{qx} and \Eq{qt} arise in the geometry of deterministic fractals, i.e., sets described by some maps with fixed parameters. When the maps are defined on the real domain, these sets are totally disconnected and characterized by a discrete scale invariance. In the case of Eqs.\ \Eq{qx} and \Eq{qt}, this scale invariance is $F_\om(\la x^\mu)=F_\om(x^\mu)$, where $\la=\exp(-2\pi/\om_N)$. Thus, multifractional spacetimes described by the measure \Eq{qx}--\Eq{qt} exhibit a fundamentally discrete geometry at scales near $\lp$ and a multiscale coarse-grained continuous geometry at scales $\sim \ell_*$ \cite{frc2}. The spacetime thus defined has a number of characteristic features including a scale-dependent dimension and a cyclic early-universe cosmology \cite{frc11}. 

If we demand dimensional flow in momentum space rather than (or together with) position space, the second flow-equation theorem establishes the profile of the spectral dimension $\ds$ at mesoscopic-to-large scales and a unique asymptotic form of the return probability (we will not use any of these concepts later).

Since the measure is neither translation nor Lorentz invariant, all Poincaré symmetries are broken in the ultraviolet but are recovered in the infrared. This situation, not uncommon in many bottom-up models of quantum gravity, typically requires the choice of a frame where physical observables are computed. Such a frame is part of the definition of multifractional theories and is called fractional picture. Its properties are an interesting chapter of the paradigm which, however, we will not examine in detail here; for a full discussion, see \cite{trtls} and the update \cite{revmu}. The bottom line is that, when observables are computed carefully, no inconsistency arises in the theory, not even at the quantum level where Lorentz violations can become a serious issue in traditional Lorentz-breaking extensions of the Standard Model \cite{revmu}. Operationally, choosing a frame means fixing the scaling of the fractional coordinates $x^\mu$ so that it is constant ($[x^\mu]=-1$), while the scaling of the variable part of the geometric coordinates $q^\mu(x^\mu)$ is scale-dependent and anomalous ($[q^\mu]\sim [|x^\mu|^{\a_\mu}]=-\a_\mu$ in the ultraviolet). Physically, the frame and unit choice consists in establishing that our measurement devices do not adapt with the observation scale and observations at different scales require different apparatus. This prescription describes an observer living in a multiscale spacetime.

A multiscale or multifractional spacetime geometry can also be multifractal, provided the Hausdorff, spectral and walk dimensions ($\dh$, $\ds$, $\dw$) obey the relations \cite{trtls}
\be\label{proC}
\dw=2\frac{\dh}{\ds}\,,\qquad \ds\leq\dh\,.
\ee
These relations belong to the two sole contact points we can touch, in the context of quantum gravity, with the traditional descriptive definition of spatial multifractal sets \cite{trtls} (the other aspect is nowhere differentiability). In order to check \Eq{proC}, one must define the dynamics. The general framework includes three inequivalent theories (plus a toy model with ordinary derivatives), which have the same measure as above and differ only for the symmetries of the Lagrangian. The latter can have three types of derivative operators: weighted derivatives, $q$-derivatives and fractional derivatives. In this paper, we will be interested in studying the dispersion relation $E^2=E^2(k)$ in two of the three extant multifractional theories. For the theory with weighted derivatives \cite{frc2,frc11}, the dispersion relation is the usual one $E^2=k^2+m^2$ for all massive particles \cite{revmu} and, to a first approximation, there is no measurable Lorentz violation in the type of experiments considered here. In that case, the electrodynamics bound \cite{frc13} remains the strongest to date. The theory with fractional derivatives is much more interesting, not only because its dispersion relations are nontrivial \cite{revmu} but also because it is a top-down candidate (i.e., from theory to experiments) for quantum gravity (the theory with weighted derivatives is not because it does not have improved renormalizability). However, this theory is difficult to deal with directly and a more convenient way to explore it is to consider the much simpler theory with $q$-derivatives, which can be regarded either as an approximation of the case with fractional derivatives \cite{revmu} or as a stand-alone exact theory. Either interpretation is fine in what follows. 


\subsection{Multifractional theory with \texorpdfstring{$q$}{}-derivatives} 

The theory with $q$-derivatives is reviewed in this subsection. We already described the integration measure and we only have to sketch the dynamics. The action for some generic degrees of freedom $\phi^a$ in flat space is
\be
S=\int\rmd^4q(x)\,\cL[\phi^a,\p_{q(x)}\phi^a]\,,
\ee
where the Lagrangian $\cL$ is defined to be the usual one (for a scalar field, for the Standard Model and so on) with the formal replacement $x^\mu\to q^\mu(x^\mu)$ everywhere. This is \emph{not} a trivial coordinate transformation because the theory is not Lorentz invariant; it is only a convenient tool to write down a much easier version of the physical-frame Lagrangian $\cL[\phi^a,(\p_x q)^{-1}\p_x\phi^a]$ (see questions {\it 24}, {\it 25} and {\it 28} of \cite{revmu}). It is part of the definition of the theory to fix a reference frame where physical observables can be evaluated and compared with experiments. This necessity stems from the fact that the underlying geometry is characterized by an explicit hierarchy of scales. Once the physical frame and the symmetries of the theory are fixed, the Lagrangian is fully determined thereon and it takes the above schematic form. That Lagrangian can be \emph{formally} recast as the simple Lorentz-invariant Lagrangian $\cL[\phi^a,\p_q\phi^a]$ but this is only a practical mathematical tool helpful to extract the observables. The two sides of the replacement $x^\mu\to q^\mu(x^\mu)$ represent the parametrization of different measuring devices, scale-independent on the left-hand side (physical devices, physical frame spanned by the coordinates $x^\mu$) and scale-dependent on the right-hand side (geometric frame spanned by the composite coordinates $q^\mu(x^\mu)$). Also in scalar-tensor theories there are two frames with different measurement units (the Jordan and the Einstein frame), and their inequivalence is determined by some physical principle assumed \emph{a priori}, for instance the requirement of respecting some energy condition or the equivalence principle. A difference with respect to multifractional theories, however, is that in our case this inequivalence holds already at the classical level, while in the scalar-tensor case one must consider the quantum theory to discriminate between the two frames.

Inclusion of gravity is not difficult \cite{frc11} and the only subtlety one must really care for from the beginning is that the metric structure is independent of the measure structure. In other words, the measure structure affects the dynamics of all fields, including the gravitational one.

The action for gravity and for the Standard Model (all summarized in \cite{revmu}) can be found in \cite{frc11,frc13}. However, no detailed dynamics is needed for our results and it is very easy to see what type of dispersion relations we find in the theory by looking just at the prototypical case of a scalar field in flat space. Then the free Lagrangian reads\footnote{Here, we use a dot to denote Einstein summation. The full expression clarifies the summation convention when the same index is repeated three or more times.}
\ba
2\cL&=&-\eta^{\mu\nu}\cdot \p_{q^\mu}\phi\p_{q^\nu}\phi-m^2\phi^2\nonumber\\
&=&-\eta^{\mu\nu}\cdot \left(\frac{1}{\p_\mu q^\mu\p_\nu q^\nu}\p_\mu\phi\p_\nu\phi\right)-m^2\phi^2\nonumber\\
&=&\frac{\dot\phi^2}{(\dot q^0)^2}-\sum_i\frac{(\p_i\phi)^2}{(\p_i q^i)^2}-m^2\phi^2\,.\nonumber
\ea
The reader should not be tricked into thinking that the only modification of the dynamics amounts to some factors in front of the kinetic and gradient terms. First, these factors are not a simple conformal factor $\varOmega^2 \eta_{\mu\nu}$ in front of the Minkowski metric, which would be the same in front of all the derivative terms. Second, the factors in the action are not arbitrary and have a precise functional form (Eqs.\ \Eq{qx} and \Eq{qt}) dictated by the second flow-equation theorem or, equivalently, by fractal geometry. Since this new structure is nondynamical and independent of the metric structure, it does not correspond to a nonminimal coupling with some extra degrees of freedom. Third, when looking at the predictions of more sophisticated systems such as the multifractional Standard Model or general relativity, it becomes clear that the multiscale geometry heavily affects virtually all sectors of physics. The rigidity of the measure and its endemic influence on the dynamics are the two main reasons why these theories are easily falsifiable. This paper will demonstrate just that.

Coming back to the question about the relations \Eq{proC}, one can prove that they are indeed satisfied in the theory with $q$-derivatives \cite{revmu}. Therefore, these spacetimes are not only multifractional but also multifractal. The same holds for the theory with fractional derivatives, while the theory with weighted derivatives does not describe a multifractal spacetime.


\subsection{Multifractional dispersion relation} 

In compact notation, the classical equation of motion for the massive scalar field is $(\p_{q^\mu}\p^{q^\mu}-m^2)\phi=0$, which we now rewrite in momentum space. The theory with $q$-derivatives admits a unitary and invertible Fourier transform mapping position to momentum space \cite{frc11}. In four topological dimensions, and independently of the specific form of the profiles $q^\mu(x^\mu)$,
\be
\phi_p(k):=\int_{-\infty}^{+\infty}\frac{\rmd^4 p(k)}{(2\pi)^2}\,\rme^{\rmi p_\mu(k^\mu) \cdot q^\mu(x^\mu)}\phi(x)\,,
\ee
where $\rmd^4 p(k)=\rmd p^0(E)\,\rmd p^1(k^1)\cdots \rmd p^3(k^3)$ and the composite momenta 
\be
p^\mu(k^\mu):= \frac{1}{q^\mu(1/k^\mu)}
\ee
(with position-space scales $t_*,\,\ell_*$ replaced by the inverse of energy-momentum scales $E_*,\,k_*$) are by definition conjugate to $q^\mu(x^\mu)$. Then it is straightforward to recast the equation of motion as $(p_\mu p^\mu+m^2)\phi_p=0$, which yields the massive dispersion relation (here $E=k^0$)
\be
[p^0(E)]^2=|{\bf p}|^2+m^2=\sum_i [p^i(k^i)]^2+m^2\,.
\ee
A similar inspection of the linearized gravitational action gives the dispersion relation for the graviton. This dispersion relation replicates the pole structure of the rest of the particles of the theory \cite{revmu}, and in what follows it is enough to obtain the main results, which do not depend on the specific tensorial structure of propagators. Also, curvature effects are negligible and, as invariably done in the literature of modified dispersion relations, we can ignore the impact of the classical gravitational background.

The review ends here. From now on, $m^2=0$ and we work with the measure \Eq{qx}--\Eq{qt} and the conjugate $p^\mu(k^\mu)$. For small fractal corrections, we can write the spatial part as
\be\label{bfporig}
|{\bf p}|^2\simeq \sum_ i k_i^2\left[1-\frac{2}{\a}\left|\frac{k_i}{k_*}\right|^{1-\a} F_\om(k_i)\right].
\ee
Let us pause for a moment and discuss an interesting caveat about momentum space. We need to perform an approximation of \Eq{bfporig} in order to have a simple expression in terms of the absolute value $k$ of the momentum, rather than of its three directional components $k^i$. This approximation can be done in different ways, all of which must give very similar results since the corrections to the standard dispersion relation are small. For instance, taking the average of spatial momentum (or if the signal is nearly isotropic), one has $|k_i|\simeq k/\sqrt{3}$ and, defining $K_*=\sqrt{3} k_*$, we get
\be\label{bfp1}
|{\bf p}|^2\simeq k^2-\frac{2K_*^2}{\a}\left(\frac{k}{K_*}\right)^{3-\a} F_\om\left(\frac{k}{\sqrt{3}}\right)\,.
\ee
Alternatively, choosing a frame where $p_i(k_i)=(p(k),0,0)_i$, one has
\be\label{bfp2}
|{\bf p}|^2\simeq k^2-\frac{2k_*^2}{\a}\left(\frac{k}{k_*}\right)^{3-\a} F_\om(k)\,.
\ee
Although this is the same as picking a frame $k_i=(k,0,0)_i$, momentum space is not Lorentz invariant in the usual way. The theory is invariant under the nonlinear transformations $p^\mu({k'}^\mu)=\Lambda_\nu^{\ \mu} p^\nu(k^\nu)$ (discussed in position space as $q^\mu({x'}^\mu)=\Lambda_\nu^{\ \mu}q^\nu(x^\nu)$ in, e.g., \cite{frc13,trtls}), but this is not a symmetry in the frame where predictions are made (the so-called fractional picture). It is easy to see why. By the very definition of the theory, in the physical frame one works with position coordinates $x^\mu$ and momentum coordinates $k^\mu$, which is the same as to use clocks and rods that do not change with the scale of observation. Then the scales in the measure appear explicitly in the formul\ae\ and break ordinary Lorentz invariance. On the other hand, the theory is formally invariant under $q$-Lorentz transformations but these do change the physics, as shown by Eqs.\ \Eq{bfp1} and \Eq{bfp2}: the characteristic scale changes from $K_*$ to $k_*$. This situation is very similar to the known problem of presentation of the measure \cite{trtls}: one must choose the coordinate frame $\{k^\mu\}$ in which the above profile $p^\mu(k^\mu)$ is defined.\footnote{In position space, the problem of presentation can be stated as follows. The theory with $q$-derivatives breaks Poincaré invariance explicitly and one must choose a coordinate frame $\{x^\mu\}$ where to define the profile \Eq{qx}--\Eq{qt}. This choice of frame is part of the definition of the theory and different frames correspond to different theories. For this reason, observing experimentally presentation effects would not imply any internal inconsistency in the framework. To put it in other words, the problem of presentation is very similar to the well-known It\^o--Stratonovich dilemma in stochastic mechanics \cite{trtls}, where integration of a nowhere-differentiable Wiener process can be defined with two major different prescriptions. Both prescriptions are valid but not simultaneously: simply, they describe systems with different stochastic properties. For a detailed discussion of frame and presentation dependence of physical observables, see \cite{frc13,trtls}.} In the case of GWs, we make the isotropic choice \Eq{bfp1}, for a reason we will explain in the next paragraph. 

The correction in Eq.\ \Eq{bfp1} is negative definite only if the oscillatory contribution in $F_\om$ is positive definite. Combining this equation with $[p^0(E)]^2\simeq E^2-(2E_*^2/{\a_0})({E}/{E_*})^{3-\a_0} F_\om(E)$ (we assume $E\geq 0$), taking the approximation $E\simeq k$ (consistent with assuming that corrections are subdominant), and identifying the energy scale $E_*$ with the inverse of the time and length scales $t_*$ and $\ell_*=1/k_*$ in Planck units \cite{frc13,frc12},
 we get the full dispersion relation
\ba
E^2&\simeq& k^2+2E_*^2\left[\frac{1}{\a_0}\left(\frac{k}{E_*}\right)^{3-\a_0} F_\om(k)\right.\nonumber\\
&&\qquad\qquad\left.-\frac{3}{\a}\left(\frac{k}{\sqrt{3} E_*}\right)^{3-\a} F_\om\left(\frac{k}{\sqrt{3}}\right)\right].\label{dire1}
\ea
For the $k_i=(k,0,0)_i$ choice, the factors $\sqrt{3}$ disappear and the net effect is zero for $\a=\a_0$. As we will see later, this fact may lead to a crucial restriction of the parameter space to avoid the strongest experimental bounds, but we should interpret it with care. The choices $|k_1|=|k_2|=|k_3|\simeq k/\sqrt{3}$ and $k_i=(k,0,0)_i$ look like, but are not, different presentations of the momentum measure. In fact, we fixed the presentation in position space in Eqs.\ \Eq{qx} and \Eq{qt} and, by conjugacy of position and momentum space, Eq.\ \Eq{bfporig} is a consequence of that choice. On the other hand, Eqs.\ \Eq{bfp1} and \Eq{bfp2} stem from slightly inequivalent approximations of \Eq{bfporig} which, however, should give the same phenomenology because of their resemblance with a presentation choice.\footnote{Let us expand this statement. Some of the results established in \cite{frc1,trtls} limit and refine the consequences and scope of inequivalent presentations. In particular, the qualitative features of the theory are not affected by a change in presentation because the latter leaves the anomalous scaling of the geometry unaltered. Since all new effects arise from dimensional flow and the latter is not deformed greatly, their characteristics may be presentation-independent. Whether this is true or not depends on the details of the observation or experiment. While there exist ideal examples where different presentations can be discriminated by experiments \cite{trtls}, in all concrete cases examined until today presentation effects turn out to be smaller than the accuracy of the observational constraints on the parameters of the theory \cite{frc12,frc13}. In the case of the present paper, if we put the approximations \Eq{bfp1} and \Eq{bfp2} on equal footing with a presentation choice (thus temporarily ignoring the fact that they come from the same presentation in position space), then we can expect to get similar experimental constraints.} This is actually true (see Sect.\ \ref{disc}) except in the case $\a=\a_0$, when a cancellation happens in the $k_i=(k,0,0)_i$ analog of Eq.~\Eq{dire1} and the net effect of anomalous geometry is zero. In this way, one would avoid all the constraints found below. However, we regard this cancellation as accidental, both because it stems from an approximation rather than an actual presentation effect and because we do not see any similar phenomenon in other experiments where such approximation is not made.

In what follows, we consider two cases according to the classification of \cite{trtls}. (a) \emph{Time-like fractal geometries} (trivial measure in spatial directions) with ${\bf p}={\bf k}$ and averaged or no log oscillations, $F_\om=1$. The averaging procedure \cite{frc2} is a coarse graining of spacetime and momentum geometry, which simply amounts to considering energy scales smaller than the Planck mass $\mpl$ (the scale at the bottom of the hierarchy and governing log oscillations) but larger than $E_*$. The dispersion relation \Eq{dire1} simplifies to
\be\label{dire2}
E^2\simeq k^2+\frac{2E_*^2}{\a_0}\left(\frac{k}{E_*}\right)^{3-\a_0}
\ee
and the correction is positive definite. (b) \emph{Space-like fractal geometries} (trivial measure in the time-energy direction) with $p^0=E$ and averaged or no log oscillations, $F_\om=1$. The dispersion relation $E^2\simeq E_{\rm full}^2(k)$ 
 becomes
\be\label{dire3}
E^2\simeq k^2-\frac{2E_*^2}{3^{\frac{1-\a}{2}}\a}\left(\frac{k}{E_*}\right)^{3-\a}
\ee
and the correction is negative definite.

Generic configurations with fractional time and space directions can produce corrections of either sign, periodically suppressed by the log oscillations. Cases (a) and (b) are extreme representatives of this spectrum of possibilities, both corresponding to corrections with a unique sign and maximal amplitude. Using the definition \Eq{gruve}, differentiating Eqs.\ \Eq{dire2} and \Eq{dire3} on both sides, and replacing $k\to E$ consistently with the small-mass small-correction approximation, we get\footnote{Here we are comparing the multiscale correction with a standard dispersion relation but this is not entirely correct, since also photons are affected. Taking this into account leads to an $O(1)$ correction of Eq.\ \Eq{mufef} which, however, does not alter the numerical bounds \cite{revmu}.}
\bs\label{mufef}\ba
\varDelta v_+ &\simeq& \frac{3-\a_0}{\a_0}\left(\frac{E}{E_*}\right)^{\!1-\a_0},\\
\varDelta v_- &\simeq& -\frac{3-\a}{3^{\frac{1-\a}{2}}\a}\left(\frac{E}{E_*}\right)^{\!1-\a}.
\ea\es


\section{Results}\label{resu}

\begin{enumerate}
\item[(i)] Since $0<\a_0,\a<1$, then $\varDelta v_+>0$ and $\varDelta v_-<0$. This is in agreement with the findings of \cite{trtls}, where it was shown that relativistic or nonrelativistic bodies move faster (slower) in geometries with time-like (respectively, space-like) fractal directions. Here we extend this conclusion to the group velocity of propagating waves. This is the first main result of the paper.
\item[(ii)] The corrections in Eq.\ \Eq{mufef} are less suppressed than those in Eqs.\ \Eq{devqg} and \Eq{devbh}. Although a comparison with some constraints on quantum-gravity scales was made in \cite{frc13}, this is the first time that multifractional spacetimes are directly compared with quantum-gravity models on a specific observable. It turns out that the multiscale effect is, in general, much stronger and more sensitive to observational constraints. Inverting Eq.\ \Eq{mufef}, one has
\ba
E_* &=&\left(\frac{\a_0}{3-\a_0}\varDelta v_+\right)^{-\frac{1}{1-\a_0}}E\qquad \text{(time-like)}\,,\label{tlest}\\
E_* &=&\left(\frac{3^{\frac{1-\a}{2}}\a}{3-\a}|\varDelta v_-|\right)^{-\frac{1}{1-\a}}E\qquad \text{(space-like)}\,.\label{slest}
\ea
We extract two types of bounds, an ``absolute'' one and one for a specific choice of $\a_0$ or $\a$. In the first case, using Eq.\ \Eq{devobs} and plotting, for instance, Eq.~\Eq{tlest} as a function of $\a_0$, one finds the conservative lower bound (T) $E_*> 15\,{\rm MeV}$ at $\a_0\approx 0.02$. This value of $\a_0$ has no particular significance theoretically, just like similar ones found in \cite{frc13,frc12}. The parameter $\a_0$ is free in the range $(0,1)$ and the justification to take small values $\a_0\ll 1/2$ is simply to find the weakest possible bounds on the theory starting from $\a_0$-dependent equations such as \Eq{tlest} (the same discussion holds for $\a$ and the space-like-fractal case). These bounds (called ``absolute'' in \cite{frc13,frc12}) set the lowest possible energy scales admitted by experiments and they represent the most conservative scenario when pitting the theory against observations.

On the other hand, theoretical arguments \cite{frc1} may select $\a_0=1/2$ at the center of the interval $(0,1)$ as somewhat preferred; it also provides a concrete example of the typical size of the corrections. For this central value, we get a tremendous boost to energies of the order of (${\rm T}'$) $E_*^{(\a_0=1/2)}> 5.7\times 10^{14}\,{\rm GeV}$. For a space-like-fractal spacetime, Eq.\ \Eq{slest}, we have instead the absolute bound (S) $E_* > 8.5\,{\rm MeV}$ (at $\a\approx 0.02$), and the $\a=1/2$ bound (${\rm S}'$) $E_*^{(\a=1/2)} > 3.3\times 10^{14}\,{\rm GeV}$. Translating these constraints to bounds on the characteristic time scale $t_*$ and length scale $\ell_*$ of the geometry, we find the numbers reported in Table \ref{tab1}.\footnote{After the submission of this paper, a work appeared placing constraints on the quantum-gravity mass scale appearing in a modified dispersion relation for the graviton \cite{YYP}. The Fisher analysis therein is based on frequencies $f=\om/(2\pi)=100\,{\rm Hz}$, corresponding to $\om\approx 630\,{\rm Hz}\approx 4.1\times 10^{-13}\,{\rm eV}$ and $|\varDelta v|<4.2\times 10^{-20}$. Using these numbers, the constraints in Table \ref{tab1} from gravitational waves are strengthened by $2-4$ orders of magnitude: for instance, $E_*^{(\a_0\ll 1/2)}>4\,{\rm GeV}$ and $E_*^{(\a_0=1/2)}>5.9\times 10^{18}\,{\rm GeV}\gtrsim 0.1\,\mpl$, thus pushing the $\a_0=1/2$ model further to its limit.} The conversion requires the use of the Planck time, length, and mass, justified in \cite{frc13,frc12} using the results of \cite{ACOS}.
\begin{table*}[t]
\caption{Absolute and $\a_0,\a=1/2$ bounds on the scale hierarchy of the theory with $q$-derivatives without log oscillations. Energy bounds are obtained directly from GWs or GRBs without any unit conversion. All figures are rounded and ``$\sim$'' indicates crude estimates.\label{tab1}}
\begin{center}
\begin{tabular}{lcccc}\hline
Bounds  & $t_*$ (s)   & $\ell_*$ (m) & $E_*$ (GeV) & \\\hline\hline
GW ($\a_0,\a\ll \tfrac12$) & $<10^{-22}$  & $<10^{-14}$  & ${>10^{-2}}$ & ${\rm (T),(S)}$ \\
GRB ($\a_0,\a\ll \tfrac12$) $\sim$     & $<10^{-32}$ & $<10^{-24}$ & ${>10^{17}}$ & (TS) \\
GW ($\a_0,\a=\tfrac12$) & $<10^{-39}$    & $<10^{-30}$ & ${>10^{14}}$ & ${\rm (T'),(S')}$ \\
GRB ($\a_0,\a=\tfrac12$) $\sim$ & $<10^{-50}$ & $<10^{-42}$ & ${>10^{35}}$ & ${\rm (TS')}$ \\\hline
\end{tabular}
\end{center}
\end{table*}
\item[(iii)] If we compare these numbers with those of \cite{frc13}, we see that the absolute bounds (T) and (S) are very close to the lower limit $E_*> 10\,{\rm MeV}$ found in the Lamb-shift effect. Thus, we have shown that the naive expectation that ``astrophysical constraints are stronger than Earth-based constraints from precision experiments'' is valid for this theory only if we fix the fractional exponents to $O(0.5)$ values. If we let them free, then GW experimental bounds may be of the same order as the Lamb-shift bound.
\item[(iv)] On the other hand, the bounds (${\rm T}'$) and (${\rm S}'$) are 12 orders of magnitude stronger that the Lamb-shift bound for $\a_0=1/2$. Thus, we have improved the constraints of \cite{frc13,frc12} for the $q$-theory and found, for the first time, bounds coming from the spatial directions.
\item[(v)] Let us now study the role of log oscillations. Considering a nontrivial log-periodic profile $F_\om\neq 1$, the bounds (T), (${\rm T}'$), (S), and (${\rm S}'$) become weaker due to a modulation effect. For instance, using Eq.\ \Eq{dire1} for a time-like fractal, Eq.\ \Eq{tlest} is replaced by 
\ba
E_*^{{\rm (log)}}\! &=&\! \left\{(\a_0|\varDelta v_+|)^{-1}\!\left[3-\a_0+\!A'\cos\!\left(\!\om_N\ln\frac{E}{\mpl}\right)\right.\right.\nonumber\\
&&\left.\left.+B'\sin\left(\om_N\ln\frac{E}{\mpl}\right)\right]\right\}^{\frac{1}{1-\a_0}}E\,,\label{tlestlog}
\ea
 where $A'=(3-\a_0)A+B\om_N$ and $B'=(3-\a_0)B-A\om_N$. To deal with the oscillatory part, we notice that we have two free parameters ($A$ and $B$) and one free but discretized parameter $\om_N\approx 4.53,2.86,\dots$. Fixing $\a_0=1/2$ and picking the first few values of $\om_N$, we have checked that log oscillations do not change the bound (${\rm T}'$) by more than one order of magnitude for $0<A,B<1$, the range that guarantees that the measure is positive definite (Fig.\ \ref{fig1}).
 However, for specific choices of $A$ and $B$ the ratio $E_*^{{\rm (log)}}/E_*$ can drop down to nearly zero, meaning that the scale $E_*^{{\rm (log)}}$ becomes virtually unconstrained. Taking, for example, $N=2$ and $A=0$, we get the minimum $E_*^{{\rm (log)}}/E_*\sim 10^{-32}$ at $B\approx 0.676505$, while $E_*^{{\rm (log)}}/E_*\sim 10^{-12}$ for $B\approx 0.676504$. A generic drop of one order of magnitude, $E_*^{{\rm (log)}}/E_*<0.1$, occurs for $0.5<B<0.9$. We conclude that, in order to milden the lower bound on the energy one must fine tune $A$ and $B$ to at least one part over ten, while to avoid a strong bound altogether (say, $E_*^{{\rm (log)}}>1\,{\rm TeV}$) the fine tuning is of at least one part over $10^7$. This the fifth result of the paper.
\begin{figure}
\centering
\includegraphics[width=8cm]{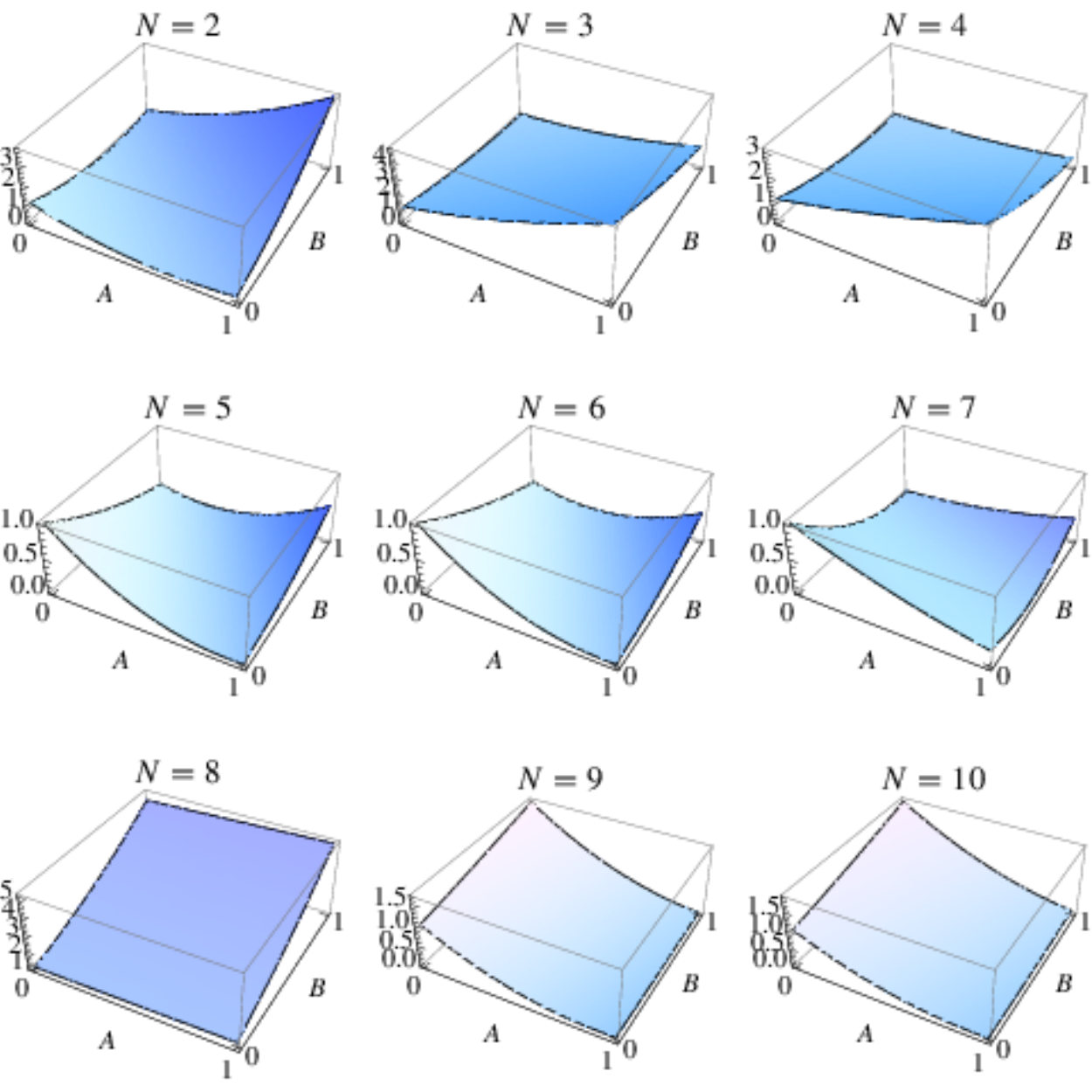}
\caption{\label{fig1} The ratio $E_*^{{\rm (log)}}/E_*$ (Eqs.\ \Eq{tlestlog} and \Eq{tlest}) as a function of the amplitudes $A$ and $B$, for $N=2,3,\dots,10$.}
\end{figure}
\item[(vi)] The constraints (T), (${\rm T}'$), (S), and (${\rm S}'$) can easily be improved by accounting for photon time delays in highly energetic events such as gamma-ray bursts. We do not present a detailed calculation of the effect in the theory with $q$-derivatives, as it is not necessary: a very crude estimate will suffice for our purpose. Let us write down Eq.\ \Eq{mufef} as $v=1+\g (E/E_*)^{1-\b}$, where $\pm\g=O(1)-O(10)$ and $\b=\a_0,\a$. The difference in the velocities of two photons with different energies emitted in a GRB at the same time is $\de v=v_2-v_1=\g (E_2^{1-\b}-E_1^{1-\b})/E_*^{1-\b}$. Taking $E_2\gg E_1$ (highly energetic photons), one gets $E_*\sim E_2/\de v^{1/(1-\b)}$. Letting $d$ be the luminosity distance between the source and us and $\varDelta t=t_2-t_1$ the time delay in the arrival of the photons, we also have $1\gg \de v\sim d/t_2-d/t_1\simeq d\varDelta t/t_1^2\simeq v_1^2\varDelta t/d\sim \varDelta t/d$. The observed sources of bright GRBs are in the range of redshift $z=0.16-3.37$ (i.e., \cite{BAJPP}), corresponding to $d\sim 10^{25}-10^{27}\,{\rm m}$. For typical photon emissions, $\varDelta t\sim 10^{-2}-10^{-1}\,{\rm s}$, so that $\de v\sim 10^{-20}-10^{-18}$. Taking $E_2\sim 100\,{\rm keV}$, we get
\be\label{estilas}
E_*>E_*^{\rm max}=10^{-1+18/(1-\b)}-10^{-1+20/(1-\b)}\,{\rm GeV}\,.
\ee
For $1-\b=2$ and $1-\b=1$, we have $E_*>10^8-10^9\,{\rm GeV}$ and $E_*>10^{17}-10^{19}\,{\rm GeV}$, in agreement with the actual estimates for, respectively, $M_2$ and $M_1$ quoted at the beginning of the paper. The $\b=0$ case also provides an absolute upper bound (TS) for the $q$-theory (Table \ref{tab1}). However, for $1-\b=1/2$, the lower bound becomes astronomically high, $E_*^{(\a_0=1/2)}>10^{35}-10^{39}\,{\rm GeV}$ (we label it (${\rm TS}'$)). Even discounting, conservatively, a few orders of magnitude with respect to a rigorous estimate, the fundamental energy scale $E_*$ would be more than 10 orders of magnitude larger than the Planck mass, thus completely ruling out the theory for $\a_0\geq 1/2$ or $\a\geq 1/2$.
\end{enumerate}


\section{Discussion}\label{disc}

This paper does not consist in a fit of some phenomenological dispersion relation. First, the dispersion relation \Eq{dire1} is derived rigorously from the theory with $q$-derivatives and it constitutes a top-down prediction which can be tested by experiments. Second, the form of \Eq{dire1} is unique to this theory and there is no other proposal, either top-down or bottom-up, reproducing it. The simplified versions \Eq{dire2} and \Eq{dire3} of \Eq{dire1} do look like the well-known phenomenological relations discussed in Sect.\ \ref{qgr}, but only because the correction is a power law when log oscillations are ignored. The power itself has a totally different geometric interpretation and values range with respect to other quantum-gravity-inspired dispersion relations. Log oscillations of \Eq{dire1} are turned on (as done in Sect.\ \ref{resu}, in particular in \Eq{tlestlog}). Third, the bounds obtained here are the first constraints that can rule out some versions of the multifractional theories and they answer a very legitimate question (What are bounds in the hierarchy scale of a multifractal spacetime?) that had been left open since the late 1970s \cite{Sti77,Svo87,ScM}. On top of this, they demonstrate that a specific feature of all quantum gravities, dimensional flow, can leave an observable imprint in some of its incarnations. Fourth, this is the first and only example to date of a theory or model that can be efficiently constrained by the recent GW observations alone.

To summarize, the fundamental energy scale $E_*<\mpl$ of the geometry is a free parameter bounded from above by the Planck mass \cite{frc2,frc11,ACOS}. For the approximation \Eq{bfp1} ($|k_i|\simeq k/\sqrt{3}$), the theory is observationally acceptable if the constraints in Table \ref{tab1} are respected. For values of $\a_0$ or $\a$ near or above $1/2$, either $E_*\gg\mpl$ (which would be theoretically inconsistent) or an unviable excess of Lorentz violation in GRB events is produced. Although it is likely that a rigorous estimate of exotic effects in GRB will not alter the main outcome qualitatively, there is no proof available yet of that. For this reason, the present GRB constraints on the $q$-theory might be regarded as preliminary. Nevertheless, it is worth discussing possible ways out. As far as we can see, there are three.
\begin{itemize}
\item[(a)] One is to consider fractional exponents $0<\a_0,\a<1/2$. The choice $\a_0=1/2=\a$ is strongly recommended by rigorous arguments for fractional-derivative spacetimes \cite{frc1}, but it is only a suggestion in the case of $q$-derivative spacetimes. Therefore, it can be abandoned without compromising the consistency of the theory. For instance, for $\a_0,\a\lesssim 0.1$ one has $E_*^{\rm max}\lesssim \mpl$.
\item[(b)] Another possibility is to account for logarithmic oscillations. Then an effect similar to that displayed by $E_*^{{\rm (log)}}$ can suppress the estimate \Eq{estilas} down to sub-Planckian scales. The price to pay, however, is an $O(10^{-7})$ fine tuning on the amplitudes $A$ and $B$ in the measure.
\item[(c)] The third case, whose physical interpretation is unclear, makes use of the effect of the isotropic approximation of Eq.\ \Eq{bfporig} on observations. If $\a_0\neq\a$, the difference between the inequivalent approximations \Eq{bfp1} and \Eq{bfp2} is not appreciable. Taking, in fact, Eq.~\Eq{bfp2} instead of Eq.~\Eq{bfp1}, we would end up with precisely the bounds (T) and (${\rm T}'$) also for space-like fractal geometries, instead of the very similar constraints (S) and (${\rm S}'$). However, as noticed above Eq.~\Eq{dire2}, if $\a_0=\a$ in the presentation choice \Eq{bfp2} then the correction to the dispersion relation cancels out and the massless on-shell condition is $E=k$: all the constraints found here would be avoided.
\end{itemize}
Physically, case (a) corresponds to geometries with a very small Hausdorff dimension, $0<\dh\ll 2$. As remarked below Eq.\ \Eq{slest}, at present there is no theoretical support for values in the range $0<\a_0,\a\ll 1/2$, since both $\a_0$ and $\a$ are free parameters in the range $(0,1)$. Therefore, case (a) is purely phenomenological. However, it is the most natural possibility among the three listed here, since it does not entail any severe fine tuning (the problem of case (b)) or accidental cancellations not backed-up by independent arguments (the problem of case (c)). Furthermore, independent bounds from the cosmic microwave background, obtained after the submission of this paper, point towards a very similar parameter range for the spatial exponent, $\a\lesssim 0.6$, but only in the presence of log oscillations \cite{frc14}. This result is robust against approximation choices in the momentum frame, so that it renders (c) less likely and yields credit to case (a) as the most plausible explanation, perhaps helped by some extra suppression of the corrections thanks to log oscillations. In future work, one might thus look into a hybrid direction (a)+(b), where $\a_0$ and $\a$ are small (certainly $<1/2$) and the tuning on the amplitudes $A$ and $B$ is not too severe. This the sixth and last result.

For completeness, one could lift the isotropic approximation and study a purely anisotropic dispersion relation. This might unravel new effects coming from having preferred directions in position and momentum space. However, in this paper we made a rather technical point showing that these effects are probably second-order with respect to those considered here. First, we showed that the isotropy approximation is very similar (although not completely equivalent) to a presentation choice, which amounts to fix the physical frame. Next, we recalled that different presentations may change some coefficients in the corrections in physical observables, but that they do not differ in the scaling property of the measure. Third, since the type of observational constraints considered in this and other papers \cite{frc12,frc13} relies just on this scaling, one can conclude (and verify explicitly) that different presentation choices are constrained by about the same bounds. Hence, one can expect that the isotropy approximation is subject to the same limitations as a presentation choice, which eventually means that the constraints found here are robust. Moreover, the GRB bound is so strong that, most likely, it will not be changed in the case of an anisotropic dispersion relation.

We conclude with a short remark about other constraints on Lorentz violations in quantum field theory. In many exotic theories beyond the Standard Model, it is possible that classically acceptable Lorentz violations be magnified to unacceptable levels by quantum mechanisms, either from an amplification by renormalization effects \cite{CPSUV,CPS} (controllable, in some cases, by carrying out a rigorous renormalization program \cite{ROR}) or as an infrared phenomenon in Unruh--DeWitt detectors \cite{HuLo}. Neither problem affects the multifractional theory with $q$-derivatives. The argument of \cite{CPSUV,CPS} was already discussed in \cite{frc9} and we will not repeat it here. Concerning the other, a crucial assumption made in \cite{HuLo} is that the correction function in the dispersion relation $E=k f(k/M)$ be $f<1$ at some point in the momentum $k=|{\bf k}|$. Recasting \Eq{dire2} and \Eq{dire3} with the same notation, one immediately sees that $f>1$ in the first case (time-like fractal geometries) and $f<1$ in the second case (space-like fractal geometries). However, this comparison is not sufficient to conclude that the space-like fractal case would be plagued by the infrared corrections considered in \cite{HuLo}. In fact, the calculation of the transition rate $\cF(\varOmega)$ from a state with zero energy to a state with energy $\varOmega$ can be performed in geometric coordinates and leads, formally, to the standard result $\cF(\tilde\varOmega)=-\tilde\varOmega\theta(-\tilde\varOmega)/(2\pi)$ for a massless scalar field (we do not present the explicit calculation, which is easy and follows exactly the same steps as the standard case, the only difference being that ``energies'' are composite). Here $\theta$ is the Heaviside step function and $\tilde\varOmega=p^0(\varOmega)\simeq \varOmega(1-{\rm corrections})$. For vanishing corrections, $\tilde\varOmega=\varOmega$ and $\cF$ vanishes identically for all positive $\varOmega$ (no spontaneous excitation of the detector). For nonzero small corrections, $\tilde\varOmega\leq\varOmega$ and equality holds only when $\tilde\varOmega=\varOmega=0$; therefore, also in this case there are no uncontrolled excitations. For negative $\varOmega$, a similar conclusion holds and there are no spontaneous low-energy de-excitations. In other words, $\cF(\varOmega)\to 0$ when $\varOmega\to 0^\pm$, contrary to the examples of \cite{HuLo} where $\cF$ tends to a finite value at small $\varOmega$.


{\footnotesize
\noindent {\bf Acknowledgments} The author is under a Ram\'on y Cajal contract and is supported by the I+D grant FIS2014-54800-C2-2-P. He thanks D.\ Rodríguez-Fernández for useful comments.

\noindent {\bf Open Access} This article is distributed under the terms of the Creative Commons Attribution 4.0 International License (\href{http://creativecommons.org/licenses/by/4.0/}{\cob http://creativecommons.org/licenses/by/4.0/}), which permits unrestricted use, distribution, and reproduction in any medium, provided you give appropriate credit to the original author(s) and the source, provide a link to the Creative Commons license, and indicate if changes were made.
Funded by SCOAP${}^3$.}

\end{document}